\documentclass{aa}

\usepackage{graphicx}    
\usepackage{txfonts}     
\usepackage{natbib}
\usepackage{algpseudocode}
\usepackage{amssymb} 
\usepackage{amsmath}
\usepackage{breqn}
\usepackage[toc, page]{appendix}
\usepackage{xcolor}
\bibpunct{(}{)}{;}{a}{}{,} 

\title{Shock-type inference of L1157 B2 using methanol desorption}
\author{Cedric Baijot\inst{1} \and Maria Groyne\inst{1} \and Michaël De Becker\inst{1}}
\institute{Space Sciences, Technologies and Astrophysics Research (STAR) Institute, University of Liège, Quartier Agora, 19c, Allée du 6 Août, B5c, 4000, Sart Tilman, Belgium\\
}

\date{Received January 31, 2025; accepted November 11, 2025}

\abstract
{
  Shock types of low-velocity molecular outflows are not always well constrained.
  Astrochemical comparisons are often made between low-velocity and high-velocity outflows, but without considering the question of the shock type.
}
{
  We   investigated molecular abundances of post-shock regions to determine whether strong differences between non-irradiated C-type and J-type shocks can be highlighted. One of the main application goals is to diagnose the shock type of the protostellar object L1157 B2 through the use of molecular tracers.
}
{
  We simulated grid sets of shock models with the Paris-Durham Shock code with velocities ranging from 5 to 19 km/s and low densities  from $10^2$ to $10^5$ cm$^{-3}$. We computed the desorption percentage of methanol in these simulations and estimated it at higher velocities.  We compared our results to observational measurements of L1157 B2 and with a benchmark of four already identified shocks.
}
{
  L1157 B2 has been diagnosed as a non-irradiated C-type shock, and the method showed a good applicability through the benchmark. Methanol formed in the icy mantle of grains can serve to trace the differences between shock types, at least in non-irradiated conditions. A requirement for the applicability of a species as a shock-type tracer is that it does not undergo significant enhancement or destruction, but is mainly impacted by desorption processes under shocked conditions.
}
{
  The desorption percentage of methanol is a good criterion in characterizing the shock type of L1157 B2 and should be investigated as a general method to diagnose the shock type in non-irradiated regions. We identify L1157 B2 as a non-irradiated C-type shock with velocities and densities fitting with previous studies.
}
\keywords{Astrochemistry -- Molecular processes -- Shock waves -- Methods: numerical -- ISM: molecules}

\begin{document}

\maketitle

\section{Introduction}
\label{sec:introduction}

The interplay between interstellar shocks and astrochemistry has been the focus of several studies in recent years. Various astrochemical codes allow us to simulate the effects of shocks on molecular abundances. Among them, the Paris-Durham Shock code simulates shocks through the computation of magnetohydrodynamical equations dynamically coupled with chemical reactions, but at the expense of a smaller chemical network \citep{1985MNRAS.216..775F, 1986MNRAS.220..801P, 2019A&A...622A.100G}. In contrast, codes such as UCLCHEM \citep{2017AJ....154...38H} or Nautilus \citep{2016MNRAS.459.3756R} parameterize the physical evolution of key parameters to concentrate on the chemical processes alone. With the direct computation of physical equations in the dynamical codes, numerous phenomena have been quantified and subsequently included in the parameterizations. This is the case for the ion-neutral drift \citep{1985MNRAS.216..775F} or the influence of the far-ultraviolet (FUV) radiation field \citep{2019A&A...622A.100G}. With these added processes, the behavior of different shock types has been reproduced: J-type  \citep{1987MNRAS.224..403F, 2020A&A...634A..17J}, C-type  \citep{1985MNRAS.216..775F, 2003MNRAS.343..390F, 2008A&A...482..549J, 2019ApJ...881...32B}, C$^*$-type, and CJ-type shocks \citep {2019A&A...622A.100G}. In particular, J-type and C-type shocks refer to outflows without any external irradiation in the FUV domain (hereafter referred to as non-irradiated shocks).

Despite this progress in understanding the chemistry of shocks, some questions remain. While astrochemical shock models have successfully traced the temperature, density, and velocity of shocks \citep{2011ApJ...740L...3V, 2016MNRAS.462.2203G}, the determination of the shock type often remains ambiguous. Some outflow types have not been fully characterized yet, as in the case of the protostellar object L1157 B2. The issue of whether molecular abundances can diagnose shock types, and more specifically C-type from J-type shocks, has previously been explored, yet no significant differentiation between the types has been established \citep{2020A&A...634A..17J}. Nevertheless, J-type, C-type, C$^*$-type, and CJ-type shocks show peculiar physical conditions and processes that should affect the chemistry in some specific manners.

To allow the characterization of astrophysical environments based on astrochemical observables, a basic requirement is to identify efficient molecular tracers. Molecular tracers are species that behave specifically in peculiar conditions. Among all molecules observed in space, some of them are known to be good shock tracers and indicate the presence of outflows, and thereby are able to characterize protostellar environments \citep{2021A&A...655A..65T}. These molecular tracers are deeply impacted by shocked conditions, either through enhanced formation driven by hot gas-phase chemistry or through the sputtering releasing species in the gas phase. Common examples are SiO, SiO$_2$, H$_2$CO, CH$_3$OH, NH$_3$, and H$_2$O \citep{1995ApJ...443L..37T, 2011ApJ...740L...3V, 2014MNRAS.440.1844S, 2016MNRAS.462.2203G, 2021A&A...655A..65T}. These shock tracers are studied through numerical simulations to understand their pathways, and can also serve to characterize some shock properties \citep{2016MNRAS.462.2203G, 2021MNRAS.507.1034L, 2023A&A...675A.151H}. In that context, complex organic molecules (COMs) are very interesting species. COMs are thought to mainly form in icy mantles of dust grains in quiescent molecular clouds or hot cores \citep{2009ARA&A..47..427H, 2013ApJ...765...60G}, and their release in the gas phase due to the passage of a shock can be a great source of information in such environments where ice mantles are initially present. \citet{2016ApJ...827...21B, 2019ApJ...881...32B} studied the effect of a C-type shock on COM abundances and identified distinct families of behavior: (i) species sputtered from grains; (ii) species already present in the gas phase, but enhanced by the passage of a shock; and (iii) species first sputtered from grains and then enhanced by the shock. In the results in \citet{2019ApJ...881...32B}, C-type shocks are proven to release a large part of grain material. Therefore, some COM gas phase abundances directly indicate the initial amount of COMs trapped on dust grains, provided that their gas-phase abundance before any shock is negligible  compared to the grain mantle abundance, and that it does not undergo any significant hot gas phase chemistry or solid phase chemistry before sputtering \citep{2016ApJ...827...21B, 2019ApJ...881...32B}. It is particularly helpful in constraining grain mantle abundances as gas-phase species emit in the radio regime. Nowadays, it can be coupled with \textit{James Webb} Space Telescope observations. Recent studies efficiently reported icy molecular abundances as in the case of some COMs in protostellar environments \citep{2024A&A...683A.124R}. In order to determine the shock type, we analyzed the release of one of these COMs, methanol, in non-irradiated shocks, and its dependence on the shocked conditions. In the context of protostellar outflows, methanol is a commonly observed molecule, for which the behavior in shocks has already been documented. Its abundance is enhanced in the gas phase by the passage of low-velocity shocks through the sputtering of icy mantles without increases in the grain-phase abundance \citep{2019ApJ...881...32B}, while it is simply destroyed in high-velocity shocks \citep{2014MNRAS.440.1844S}.

The main objective of this work is to assess how methanol can serve as shock-type tracers. By defining a new criterion, we propose a new method based on methanol desorption to diagnose shock types in non-irradiated environments. The characterization of non-irradiated shocks is particularly helpful in understanding dense, yet active environments, as is the case during the formation of Class 0 protostellar cores. The advantage of the method presented in this paper relies on the consideration of an observable for which the underlying physico-chemical processes present simpler dependences on shock conditions than a global comparison between models and astronomical observations. In the end, this is expected to allow  a good contrast between shock types. To describe the shock-type diagnosis through molecular tracers, the paper is organized as follows. In Sect.\,\ref{sec:L1157} we present L1157 B2, the test case where the shock type is still unknown, and that was the motivation of the current paper. Section\,\ref{sec:methods} describes the computational method and the idea behind the identification of the shock type. The characterization of shock types through methanol desorption is presented in Sect.\,\ref{sec:results} with primary interpretations, and is  discussed in Sect.\,\ref{sec:discussion}. The applicability of the method is verified on shocks for which the type is known before being applied to the test case, L1157 B2, in Sect. \ref{sec:testcase}. Finally, our summary and conclusions are  given in Sect.\,\ref{sec:conclusions}.

\section{L1157 B2}
\label{sec:L1157}

L1157 is a protostar of class 0 located at 250 pc \citep{2007ApJ...670L.131L}. It exhibits bipolar outflows, made of R and B components, and contains clumps labeled depending on their position (see Fig. 1 in \citealt{2016A&A...593L...4P}). L1157 is known to be a very good astrochemical laboratory for modelers. In the B outflow, there are two main clumps, labeled B1 and B2, both thought to harbor a shocked region. Numerous molecular abundance measurements of diverse species including methanol proved the presence of a high molecular diversity in L1157 B1 \citep{2008ApJ...681L..21A}. Studies based on NH$_3$ and H$_2$O emissions have successfully constrained the temperature, the velocity, and the density of B1 \citep{2011ApJ...740L...3V}. From these studies, B1 corresponds to a C-type shock with a shock velocity of 40 km/s and a density of $10^4$ cm$^{-3}$. In contrast, fewer observational constraints exist for B2. Its shock velocity and density match with low-velocity outflows ($\sim$ 10 km/s) in more diffuse gas ($\sim 10^3$ cm$^{-3}$) when their NH$_3$ and H$_2$O profiles are compared with models \citep{2016MNRAS.462.2203G}. The impact of the radiative field was not assessed in these studies and is therefore not constrained for either L1157 B1 or L1157 B2. However, the external FUV radiation field is thought to be negligible given that class 0 protostellar cores are situated in embedded environments, and are not themselves sources of FUV radiation. Moreover, L1157 B2 is also partially protected by L1157 B1, which beneficiates from a higher density \citep{2007ApJ...670L.131L, 2014A&A...565A..64P, 2016A&A...593L...4P}. In these studies, the shock type of L1157 B2 has never been determined. Comparisons between H$_2$O and SiO emissions with predicted values did not enable the determination of the shock type \citep{2012A&A...537A..98V}. Comparisons with the enhancement degree of gas phase species with shock models were also unsuccessful \citep{2020A&A...634A..17J}. The question of the shock type of L1157 B2 is therefore still open and constitutes our test case to evaluate the efficiency of the gas-phase release of COMs as shock-type tracer.

Our analysis is based on CH$_3$OH. The characterization of L1157 B2 was the targeted application case of our studies, and only a few different COMs have been observed in this region. In this reduced set of potential tracers, methanol is the best candidate for the current method. A discussion on possible other shock-type tracers is found in Sect. \ref{sec:Other Shock-Type tracers}. The fractional abundance of methanol in the pre-shock and shocked regions of B2 clump has been measured to be $4.5 \times 10^{-8}$ and $2.2 \times 10^{-5}$, respectively \citep{1997ApJ...487L..93B}.

\section{Methods}
\label{sec:methods}

\begin{table}[]
    \caption{Domains of definition and value of key parameters.}
    \centering
    \begin{tabular}{l l}
    \hline
        Parameter & Domain of definition \\ \hline
        Shock velocity (km/s) & [5-19] \\
        Initial density (cm$^{-3}$) & [$10^2$-$10^4$]\tablefootmark{a} \\
        Magnetic field parameter & (0.1, 1.0) \\ \hline
        Parameter & Value \\ \hline
        Cosmic ray ionization rate (s$^{-1}$) & $5.0 \times 10^{-17}$ \\
        Shape of the incident FUV field & 1 \\
        FUV radiative transfer flag & 0 \\
        $A_v$ integration flag & 0 \\
        $A_v$ to $N_H$ conversion flag & 0 \\
        Grain temperature flag & 1 \\
        CO and H$_2$O cooling flag & 0 \\
        H$_2$ FUV pumping & 0 \\
    \hline
    \end{tabular}
    \tablefoot{
    Explanations for computational parameters are available in the code documentation \citep{DocumentationShockCode}. \\
    \tablefoottext{a}{Initial density for non-irradiated shocks has been extended to $10^5$ cm$^{-3}$.}
    }
    \label{tab:parameters}
\end{table}

We ran a grid set of shock models using the Paris-Durham Shock code v1.1 \citep{1985MNRAS.216..775F, 2019A&A...622A.100G}. This set contains J-type and C-type shock conditions encompassing the approximate velocity and density found by \cite{2016MNRAS.462.2203G}. We assumed no external FUV radiation field as we focused on shocks arising from class 0 protostellar cores. Further details on the parameter space covered by our model grid are given in \ref{tab:parameters}. The magnetic field was parameterized with the prescription from \cite{1983ApJ...264..485D}: $b = 0.1$ for J-type shocks and $b = 1.0$ for C-type shocks. The density and velocity ranges were limited to $10^5$ particles per cm$^{-3}$ and 19 km/s. At higher velocities and densities, grain-grain interactions, such as vaporization, shattering, and coagulation, become non-negligible in J-type shocks \citep{2010A&A...523A..29D, 2011A&A...527A.123G, 2019A&A...622A.100G}. However, these processes are not included in the Paris-Durham Shock code despite their high impact on the chemistry of COMs in molecular shocks. For this reason, we did not extend our parameter space, but we kept it in accordance with previous values obtained for L1157 B2.

The chemical network was the one provided with the Paris-Durham Shock code v1.1. It contains 141 chemical species for 3270 reactions. Reversed reactions were included up to an endothermicity of 2\,eV. Chemical species were divided into three phases: the gas phase, the icy mantle phase, and the core grain phase. The icy mantle was therefore constitued of one single phase without any differentiation between the enveloppe and bulk of the ice. Molecules could absorb and desorb from the icy mantle phase by thermal and non-thermal (sputtering and cosmic-ray induced) processes. Icy mantle reactions were approximated by the direct saturation of adsorbed species. Many COMs were thus not considered by the Paris Durham Shock code. This limited the diversity reachable by the chemical network, but is sufficient for our analysis, as justified in the following.

\subsection{Identification of the shock type}\label{Meth:Identification}

The identification of the shock type was determined by the shock model that best matches the observations. A comparison between a complete set of gas phase species such as the one done in \cite{2020A&A...634A..17J} does not seem to work due to the high variability in the processes involved. Differences in the gas phase between J-type and C-type shocks (temperature, length scale, ion-neutral drift) act in too opposite ways to give a clear differentiation in molecular abundance between the shock types at a global level. To compare observations with shock models, we therefore made use of molecules that are particularly impacted by shocked environments, such as molecules that are dominantly formed on dust grains. Among the molecules treated by the icy mantle phase in the Paris Durham Shock code, only H$_2$S, H$_2$CO, and CH$_3$OH abundances have been measured in L1157 B2. However, H$_2$S has been disregarded as its desorption is not correctly reproduced by the code (see Sect. \ref{sec:discussion}), leading to strong divergences with reality. Only H$_2$CO, and CH$_3$OH have been considered for the downstream methodological choice.

COMs, which are mainly formed in icy mantles, are enhanced in the gas phase in C-type shocks by desorption processes such as sputtering, as reported in \cite{2019ApJ...881...32B}. The same authors have described a class of molecules that does not undergo enhanced formation or destruction in the gas and in the ice phase before sputtering. They also described molecules for which the gas phase abundance at the sputtering peak is representative of the ice phase abundance before sputtering. Finally, they identified a class of molecules for which the hot gas-phase chemistry does not significantly impact the gas-phase abundance of the desorbed species. These properties characterize molecules for which the enhancement in the gas phase can be simplified to a function of the pre-shock ice density. Between H$_2$CO and CH$_3$OH, only methanol belongs to all these classes Its behavior before the sputtering is verified in \cite{2025arXiv250922203H} and the negligible hot gas-phase pathways are confirmed in \cite{2006FaDi..133..177G}. It thus served as an indicator for this study.

\begin{table}[]
    \centering
    \caption{Methanol abundance in shock models.}
    \begin{tabular}{l l l l}
    \hline
        Shock model & CH$_3$OH$_0$ & CH$_3$OH$^*_0$ & CH$_3$OH$_{ps}$ \\ \hline
        J(5 km/s) & $3.43 \times 10^{-14}$ & $1.86 \times 10^{-5}$ & $2.06 \times 10^{-11}$ \\
        J(18 km/s) & $3.43 \times 10^{-14}$ & $1.86 \times 10^{-5}$ & $3.98 \times 10^{-10}$ \\
        C(5 km/s) & $3.43 \times 10^{-14}$ & $1.86 \times 10^{-5}$ & $2.40 \times 10^{-10}$ \\
        C(18 km/s) & $3.43 \times 10^{-14}$ & $1.86 \times 10^{-5}$ & $6.62 \times 10^{-6}$ \\ \hline
    \end{tabular}
    \label{tab:desorption}
    \tablefoot{Initial fractional abundance ($_0$) of icy-mantle phase ($^*$) and gas-phase methanol and average post-shock fractional abundance ($_{ps}$) of gas-phase methanol for different types of shocks. All shocks are evaluated at $10^3$ cm$^{-3}$.}
\end{table}

Table \ref{tab:desorption} displays the initial fractional abundance of icy-mantle phase and gas-phase methanol along with the post-shock fractional abundance of gas-phase methanol. The density of $10^3$ /cm$^3$ has been chosen to follow the initial guess of \cite{2016MNRAS.462.2203G}. An evaluation for the other densities covered by the parameter space does not change the following conclusion. By evaluating the contribution of each chemical process in our simulations, we find that the enhancement of post-shock gas-phase methanol is almost only due to desorption, which is in agreement with \cite{2006FaDi..133..177G, 2014MNRAS.440.1844S, 2025arXiv250922203H}. Gas-phase pathways leading to methanol formation exist, but require the presence of CH$_5$O$^+$ or CH$_3$OH, for which the fractional abundance is always negligible. Even in low-velocity J-type shocks where desorption is the weakest, their fractional abundance never exceed $6\%$ of the maximum fractional abundance of gas-phase methanol. The first conclusion is that desorption occurs in all shocks, regardless of their type, but the degree of desorption varies depending on whether we are simulating a J-type or a C-type shock. In the absence of gas-phase formation processes, the desorption percentage of methanol $\%_{deso}$ is written as in Eq. \ref{desorptionpercentageRAW}. Given that the initial fractional abundance of gas-phase methanol is always negligible compared to the post-shock abundance, Eq. \ref{desorptionpercentageRAW} transforms into Eq. \ref{desorptionpercentage}. To evaluate its mathematical properties,  a rigorous derivation of Eq. \ref{desorptionpercentage} is given in Appendix \ref{AppendixA}.

\begin{align}
    \%_{deso} &= \frac{\chi_{post-shock}[CH_3OH] - \chi_{pre-shock}[CH_3OH]}{\chi_{pre-shock}[CH_3OH^*]} \times 100 \% \label{desorptionpercentageRAW}\\
     &\approx \frac{\chi_{post-shock}[CH_3OH]}{\chi_{pre-shock}[CH_3OH^*]} \times 100 \% \label{desorptionpercentage}
\end{align}

\noindent where $\chi$ is the fractional abundance and $^*$ denotes the species in the icy mantle phase. The pre-shock abundance corresponds to the initial abundance, while the post-shock abundance is the averaged value between the shock front (maximum of temperature) and the end of the shock (where the temperature has reached back its initial value and where the ion-neutral drift has stopped). Table \ref{tab:desorption} justifies the passage from Eq. \ref{desorptionpercentageRAW} to Eq. \ref{desorptionpercentage} in the simulations, but this approximation can be verified in real shocked conditions, as in L1157 B2. The observed abundances of gas-phase methanol in the pre-shock and the post-shock region of L1157 B2 in \cite{1997ApJ...487L..93B} are $4.5 \times 10^{-8}$, and $2.2 \times 10^{-5}$. From these values, it is directly observable that the simplification in Eq. \ref{desorptionpercentage} holds as proven in Eq. \ref{negligible}.

\begin{align} \label{negligible}
    &\chi_{post-shock}[CH_3OH] - \chi_{pre-shock}[CH_3OH] \\ \notag
    &= 2.2 \times 10^{-5} - 4.5 \times 10^{-8} \approx 2.2 \times 10^{-5} = \chi_{post-shock}[CH_3OH]
\end{align}

\begin{figure*}
    \centering
    \includegraphics[width=1.2\linewidth]{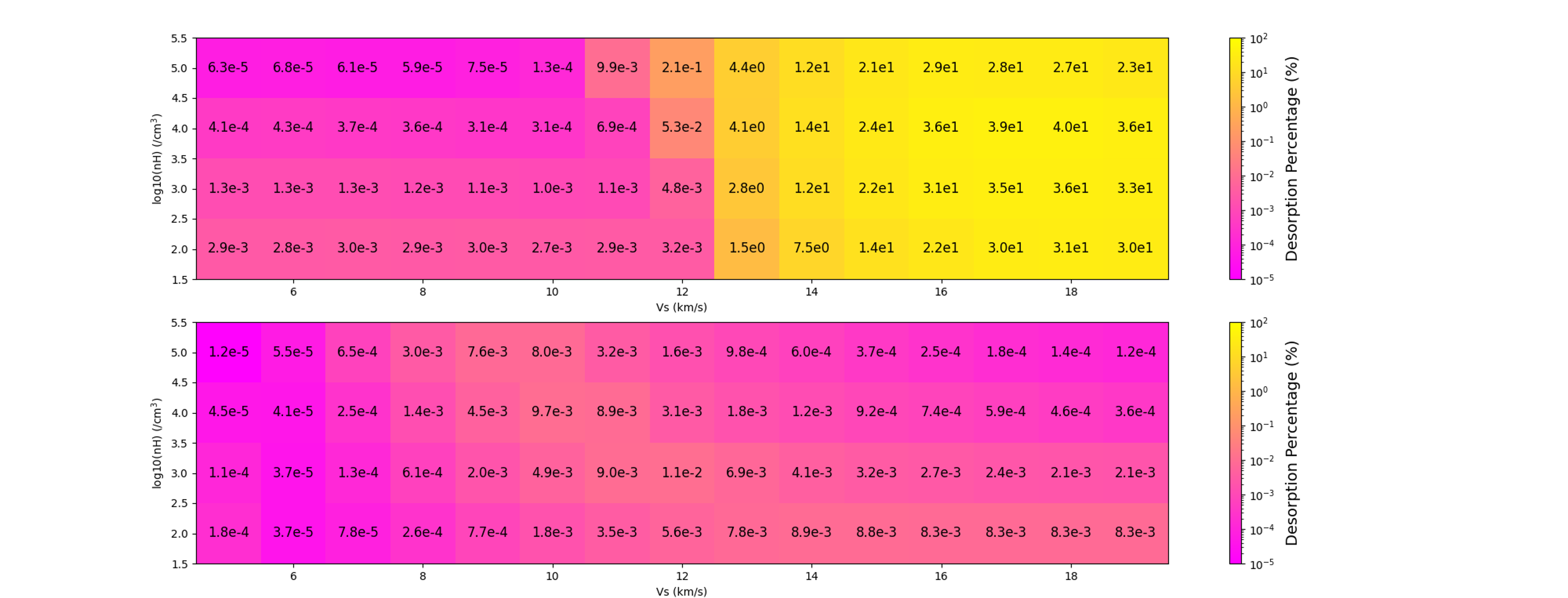}
    \caption{Desorption percentage of methanol in C-type (top) and J-type shocks (bottom). The x-axis represents the shock velocity ($V_s$) and the y-axis the pre-shock density ($nH$).}
    \label{fig:DesorptionPercentage1}
\end{figure*}

\begin{figure*}
    \centering
    \includegraphics[width=1.2\linewidth]{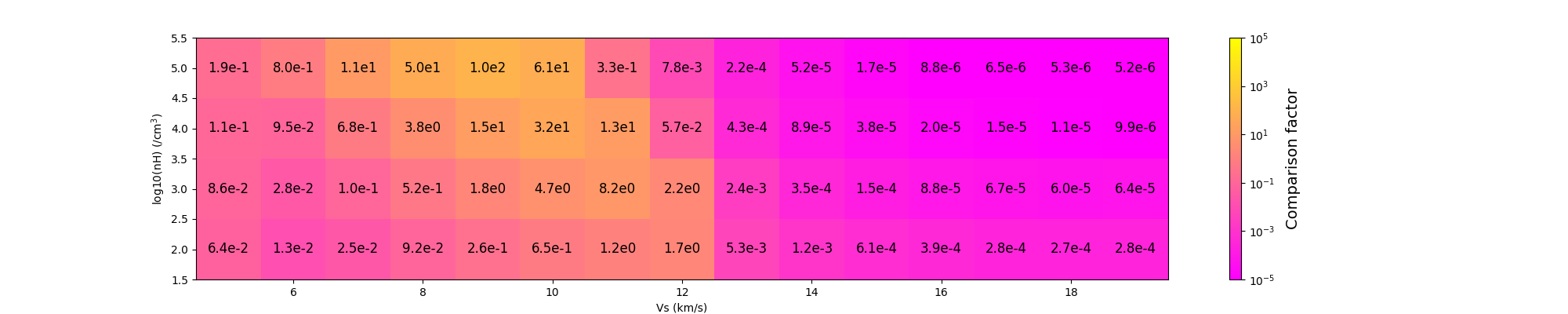}
    \caption{Comparison factor for methanol. The x-axis represents the shock velocity ($V_s$) and the y-axis the pre-shock density ($nH$).}
    \label{fig:ComparisonFactor}
\end{figure*}

In this scope, the shock model considered as the most representative of the modeled astrophysical region was considered to be the one with desorption percentage best matching the observations. By limiting our comparison to the level of desorption, we expect one benefit. The impact of shock types on desorption seems more straightforward than that of a complete set of gas-phase abundances. The only important processes impacting the desorption percentage are the ones leading to methanol desorption and methanol destruction in the gas phase. The sensitivity of the molecular abundance prediction on the chemical network is thus reduced given that several reactions will not significantly impact the results. This method is thus less model-dependent than a global comparison for a large set of species between observations and simulated values.

\subsection{Limitations of the astrochemical model}

As previously mentionned, an evaluation of the processes leading to gas-phase methanol in the simulations revealed that methanol was almost entirely provided by desorption from grain surfaces. Given that the chemical network of the Paris-Durham Shock code contains formation reactions of methanol in the gas phase, we conclude that the passage of a shock does not activate these pathways. However, the importance of a direct income of methanol from the grain phase should be discussed. Astrochemical simulations are always limited by the way to include microphysical processes, and their underlying assumptions. Even if the method decreases the impact of the chemical network by neglecting gas-phase formation processes of methanol and the impact of initial icy-mantle phase abundance by choosing a percentage of desorption (as validated in Sect. \ref{sec:Robustness}), we still identified three main limiting factors able to influence the strength of methanol desorption in the simulations:

\begin{enumerate}
    \item Chemistry on grains: The Paris-Durham Shock code approximates grain-surface chemistry by a direct saturation of adsorbed species. As the code does not possess branching ratios or any grain-surface or icy-mantle reactions, only methanol adsorption leads to icy-mantle methanol. This is unrealistic. However, as quoted in Sect. \ref{Meth:Identification} and found by \cite{2019ApJ...881...32B}, methanol does not undergo an enhancement before sputtering. The sputtering rate is therefore expected to be not impacted by the lack of grain-surface chemical reactions, but only by the initial abundance of adsorbed methanol. As this does not affect the desorption percentage, this approximation is not problematic. Nevertheless, sputtering processes are approximated by the description of the icy mantle in one unique phase. As there is no difference between the interior and the envelope of the icy mantle, the initial position of the desorbed molecule cannot be taken into account. This is a source of error as molecules only desorb from the upper monolayers according to molecular dynamics simulations performed in \cite{2010JChPh.132r4510A, 2023JChPh.159d4711A}. The position in the icy mantle is also important to correctly describe interactions with external agents such as cosmic rays or photons. Cosmic rays have a greater penetration depth while photons are limited to the upper monolayers \citep{2000A&A...357..793G, 2001JGR...10633381G}. This has an important impact on the energetic processing of the ice. More detailed calculations should be considered for greater precision in the desorption percentage.

    \item Distribution of binding energies: Several studies have proven that, in such low thermal dust grain temperature conditions, icy mantles are proven to be amorphous. Therefore, a distribution of binding energies better represents the binding behavior of chemicals \citep{Bovolenta, Ferrero, Duflot, Tinacci, 2025A&A...698A.284G}. Thermal desorption quickly becomes more important if we consider a distribution of binding energy and not a single value as shown for various adsorbates \citep{2017ApJ...839...47M}. There are thereby reasons to believe that thermal desorption processes have been underestimated in our simulations. However, the effects of thermal desorption should still have been limited. The binding energy of methanol in the Paris-Durham Shock code is 3820K \citep{2017ApJ...844...71P}. This value has been computed after visual inspection of results made by \cite{2004MNRAS.354.1133C}. In the latter study, laboratory experiments were used to determine the desorption temperature of various species by temperature-programmed desorption. The same results with different inference methods lead to a binding energy of 5534 K \citep{2017MolAs...6...22W}. Many investigations on methanol have been made and lead to a range of binding energies -- 3700 K \citep{2016A&A...585A..24M}, 3820 K \citep{2017ApJ...844...71P}, 4500 – 5100 K \citep{2017MolAs...6...22W}, 5400 K \citep{2008JChPh.128m4712B}, and 5534 K \citep{2004MNRAS.354.1133C, 2017MolAs...6...22W}. The choice of 3820 K is at the lower limit. In a single binding energy framework, choosing a smaller binding energy already increases the desorption percentage.

Recently, a binding energy distribution of methanol has been inferred in \cite{2025MNRAS.539...82B}. It approximates to a Gaussian with a mean value of 4270K and a standard deviation of 1564K ($36\%$ of the mean value). Thermal desorption studies show that the desorption percentage increases with the standard deviation up to several orders of magnitude \citep{2017ApJ...839...47M}. Nevertheless, this study has been made at a density of $10^8$ cm$^{-3}$. This is denser than the densities included in our parameter space. Density greatly impacts the value of the desorption percentage (see Fig. 4 in \citealt{2017ApJ...839...47M}). With this high density, grains have been more heated by collisions, leading to temperatures of about $\sim$ 50K. The grain temperature computed by the Paris-Durham Shock code does not exceed 30K in our simulations. Moreover, thermal desorption should have been overestimated in \cite{2017ApJ...839...47M} as they approximate the initial population of dust grain species by a Gaussian divided into $10^4$ bins. Only desorption is allowed in the grain surface processes. Binding-energy-resolved astrochemical frameworks enabling diffusion show that species tend to preferentially populate sites with high binding energies as they get trapped in them \citep{2024ApJ...974..115F}. Given that thermal desorption is mainly impacted by low-binding-energy sites (Fig. 3 in \citealt{2017ApJ...839...47M}), the inclusion of a more realistic initial population and diffusion processes should decrease the desorption percentage. Future studies are thus necessary to limit the uncertainties on the desorption percentage of methanol, leading to easier differentiation of shock models. While binding-energy-resolved astrochemical frameworks begin to appear \citep{2024ApJ...974..115F}, the computation of more reliable quantitative results based on the distribution inferred in \cite{2025MNRAS.539...82B} needs these frameworks to be suitable for highly dynamical events, such as shocks, before any applications to shock-type diagnosis.

    \item Fragmentation: Within the used chemical network, methanol that desorbs by sputtering directly joins the gas phase. However, laboratory studies \citep{2025A&A...693A..30F} have shown that molecules could have been fragmented during the process. Fragmentation decreases the post-shock gas-phase abundance of methanol given that some molecules are destroyed. It could be interesting to evaluate the impact of methanol fragmentation on the desorption percentage by multiplying the desorption percentage by the branching ratio in subsequent work. 
\end{enumerate}

\section{Results}
\label{sec:results}

\subsection{Desorption percentages of CH$_3$OH}

The desorption percentage for methanol for non-irradiated C-type and J-type shocks is displayed in Fig. \ref{fig:DesorptionPercentage1}. The range of desorption percentage strongly varies with shock conditions, going from $1.2 \times 10^{-5}\%$, up to 40\%. Comparing the two shock-type behaviors, only C-type shocks with sufficient shock velocities (Vs > 12 km/s) efficiently desorb methanol from the icy mantle phase with a percentage of desorption greater than $1 \%$. Other shock conditions display more various desorption percentages ($1.2 \times 10^{-5}$ - $2.1 \times 10^{-1}$).

\subsection{CH$_3$OH as a shock-type tracer}

Let us investigate the efficiency of CH$_3$OH to differentiate shock conditions. Figure \ref{fig:ComparisonFactor} shows the comparison factor $f_{comp}$ defined as in Eq. \ref{Eq:f_comp}:

\begin{equation}
    f_{comp} = \frac{\%_{deso}^{J}}{\%_{deso}^{C}}.
    \label{Eq:f_comp}
\end{equation}

\noindent Here $^J$ and $^C$ denote the type of shock where the percentage is evaluated. The further the comparison factor deviates from one, the better the methanol desorption behavior against given shock conditions can discriminate between shock types. As visible in Fig. \ref{fig:ComparisonFactor}, the identification of the shock type is ensured given that the shock velocity is high enough to activate an efficient desorption of methanol. Below a cutoff velocity of 12 km/s, desorption efficiency varies too much between C-type and J-type shocks due to the highly non-linear behavior of the physical conditions. At such velocities, J-type shocks and C-type shocks mainly differ by their ionic and neutral temperatures for the gas phase. Except for very well-constrained shocks, it can be difficult to identify with confidence the shock type in this domain.

The ion-neutral drift explains the barrier at a specific velocity in C-type shocks. From the work of \cite{2013A&A...550A.106L}, the velocity of $\sim 12$ km/s corresponds to an effective temperature \citep{1986MNRAS.220..801P} around the binding energy of methanol used in icy mantles (3820K in the Paris-Durham Shock code). With radiative emissions, the grains stay at relatively low temperatures \citep{2017ApJ...839...47M}, which disables thermal desorption for our range of physical conditions. In our simulations, grain temperatures never exceed 30K. An ion-neutral drift is therefore necessary to increase the effective temperature of grains and the sputtering, leading to the high difference between C-type and J-type shocks. Sputtering also occurs in J-type shocks but with reduced efficiency. The release of icy mantle materials in J-type shocks is feasible but is driven by vaporization and not sputtering \citep{2009A&A...497..145G}. As previously said, vaporization (and shattering) is not included in the Paris-Durham Shock code. However, it is not efficient at these low densities \citep{2009A&A...497..145G, 2011A&A...527A.123G, 2019A&A...622A.100G}.

\section{Discussion}
\label{sec:discussion}

\subsection{Robustness of the desorption percentage}
\label{sec:Robustness}

Our method aims to evaluate the shock type and, if possible, to better constrain shock conditions, i.e., the velocity and density, based on methanol desorption. As reduced chemical networks such as the one used by the Paris-Durham Shock code are not necessarily suitable to well reproduce initial abundance, we need a criterion not based on them. To validate this methodological requirement, we have run several shock simulations of C-type and J-type shocks at 17 km/s and with an initial density of $10^3$ cm$^{-3}$. For these simulations, the initial fractional abundance of icy mantle methanol varied from $9 \times 10^{-7}$ to $9 \times 10^{-3}$. For C-type shocks, the desorption percentage only varied from 33.7 \% to 35.7 \%. For J-type shocks, the percentage varied by about a factor of 4, from $5.7 \times 10^{-4} \%$ to $2.4 \times 10^{-3} \%$ for the most extreme cases. These variations are negligible compared to the differences between high-velocity (Vs > 12 km/s) C-type shocks and other shock conditions, with minimum discrepancies about $5 \times 10^2$ for these high velocities. The desorption percentage seems consequently robust to identify the shock type in this domain of the parameter space (i.e., in low-velocity non-irradiated C-type and J-type shocks).

\begin{figure}
    \centering
    \includegraphics[width=1\linewidth]{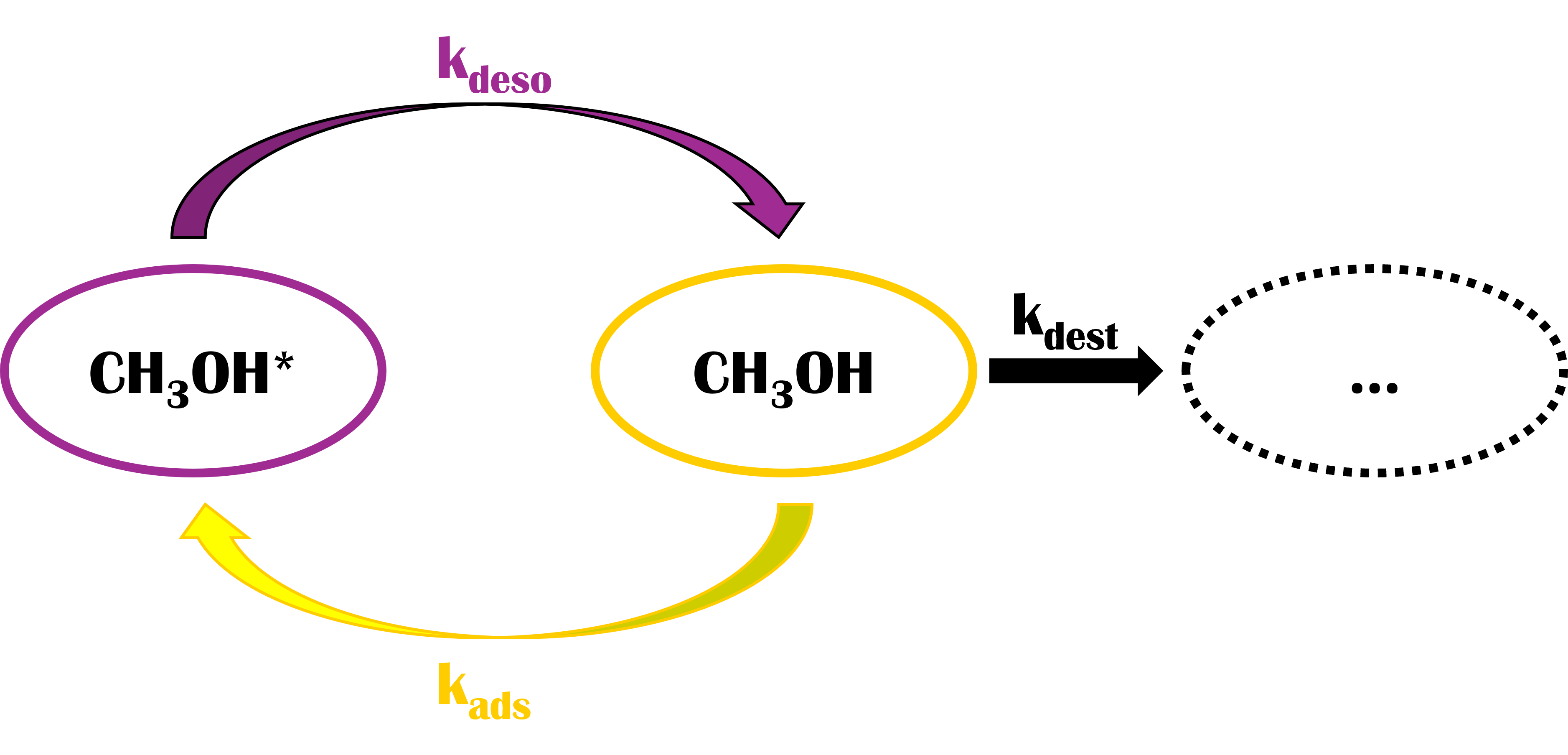}
    \caption{Reduced chemical network where gas-phase formation routes of methanol are neglected. CH$_3$OH represents the molecule in the gas phase while CH$_3$OH* is for the icy mantle phase. ``...'' represents the products of methanol destruction.}
    \label{fig:ChemicalNetwork}
\end{figure}

This stability is ensured by the fact that the post-shock abundance of gas-phase methanol is almost entirely due to the amount of icy-mantle phase methanol that has desorbed. Let's assume that gas-phase pathways leading to methanol formation can be neglected, which is justified in our simulations, and remember that no solid phase processing is undergone by CH$_3$OH. In that case, the chemical network for methanol can be reduced to methanol desorption, methanol adsorption, and methanol destruction in the gas phase (see Fig. \ref{fig:ChemicalNetwork}).

\begin{table}[]
    \centering
    \caption{Desorption reactions of icy mantle methanol.}
    \begin{tabular}{l}
    \hline
         Desorption reaction  \\ \hline
         CH$_3$OH$^*$ + photon $\rightarrow$ CH$_3$OH + grain \\
         CH$_3$OH$^*$ + secphoton\tablefootmark{a} $\rightarrow$ CH$_3$OH + grain \\
         CH$_3$OH$^*$ + crp\tablefootmark{b} $\rightarrow$ CH$_3$OH + grain \\
         CH$_3$OH$^*$ + He $\rightarrow$ CH$_3$OH + grain \\
         CH$_3$OH$^*$ + H $\rightarrow$ CH$_3$OH + grain \\
         CH$_3$OH$^*$ + H$_2$ $\rightarrow$ CH$_3$OH + grain \\
         CH$_3$OH$^*$\tablefootmark{c} $\rightarrow$ CH$_3$OH + grain \\
         \hline
    \end{tabular}
    \tablefoot{\tablefoottext{a}{Secondary photon, i.e., cosmic-ray induced photon.}\tablefoottext{b}{Reactions induced by cosmic rays.}
    \tablefoottext{c}{Thermal desorption.}
    }
    \label{tab:DesorptionReaction}
\end{table}

Table \ref{tab:DesorptionReaction} shows the relevant desorption mechanisms in our simulations. By including concentrations of He, H, and H$_2$ (assumed independent of methanol concentration) in the rate constants to form new constants $k_{deso}^i$, all reactions become first-order partial derivative equations (PDEs) in terms of [CH$_3$OH$^*$]. Photons and cosmic rays are already included in the rate constant \citep{2019A&A...622A.100G}. The desorption rate of icy mantle phase methanol can therefore be written as a global equation with a global constant $k_{deso}$ which is the sum of the $k_{deso}^i$ constants of all subprocesses. As the only reaction leading to icy mantle phase methanol is CH$_3$OH + grain $\rightarrow$ CH$_3$OH$^*$, we also write adsorption of methanol as a first-order PDE with the rate constant $k_{ads}$. Gas-phase methanol is destructed into subproducts (CH$_3$, H$_2$O, CH$_5$O$^+$, … denoted ``...'' in Fig. \ref{fig:ChemicalNetwork}) when methanol encounters with other molecules (e.g., H, H$_3$O$^+$, C$^+$). If we approximate the concentration of these reactants as being independent of methanol concentration, we can also include them in constants $k_{dest}^i$ to create a global destructive reaction associated with a constant $k_{dest}$.

The stability of the method is proven by writing the time-dependent evolution of gas-phase methanol fractional abundance $\chi[CH_3OH](t)$. With the mathematical developments given in Appendix \ref{AppendixA}, we found that it is directly proportional to the initial icy mantle phase fractional abundance $\chi[CH_3OH^*]_0$ with a proportion coefficient which is the desorption percentage from Eq. \ref{desorptionpercentage}, but at any $t$ (anywhere in the shock). Under the quasi-static approximation, the desorption percentage can be written with the system constants and takes the form (Eq. \ref{desorptionpercentagemath})

\begin{equation} \label{desorptionpercentagemath}
    \%_{deso}(t) = \frac{(k_{deso}(t) + \lambda_1(t)) \cdot (k_{deso}(t) + \lambda_2(t))}{(\lambda_2(t) - \lambda_1(t)) \cdot k_{ads}(t)} \cdot (e^{\Phi_1(t, t_0)} - e^{\Phi_2(t, t_0)}),
\end{equation}

\noindent where $\lambda_{1,2}(t)$ are given by Eq. \ref{eigenvalues}:

\begin{equation}\label{eigenvalues}
    \lambda_{1, 2}(t) = \frac{- k_{tot}(t) \pm \sqrt{k_{tot}(t)^2 - 4 \cdot k_{deso}(t) \cdot k_{dest}(t)}}{2}.
\end{equation}

\noindent Here $k_{tot}(t) = k_{deso}(t) + k_{ads}(t) + k_{dest}(t)$. $\Phi_{1,2}(t)$ are given by Eq. \ref{phi},

\begin{equation}\label{phi}
    \Phi_{1, 2}(t, t_0) = \int_{t_0}^{t} \lambda_{1, 2}(\tau) \; d\tau,
\end{equation}

\noindent where $t_0$ is the time at which the initial concentration of methanol in the ice mantle is evaluated. Based on this mathematical reasoning, the desorption percentage appears to be independent of the initial concentration of icy mantle methanol. Divergences from this result arise when we consider the system constants as being methanol concentration-dependent and when gas-phase formation pathways are included. However, gas-phase formation is negligible compared to desorption from grains especially in non-irradiated C-type shocks. This explains the robustness of the method in first approximation since the average percentage in the post-shock region is also independent of initial conditions.

\subsection{Differentiating shock types at higher velocities}

\begin{table*}[]
    \centering
    \caption{Desorption percentages of methanol in C-type shocks and J-type shocks with comparison factor.}
    \begin{tabular}{l l l l l l l l}
    \hline
        C-type shocks & Pre-shock density & Shock velocity (km/s)&  & & & & \\
         & (cm$^{-3}$) & 25 & 30 & 35 & 40 & 45 & 50 \\ \hline
         & $10^4$ & $97\%$ & $97\%$ & $96\%$ & $95\%$ & $95\%$ & $95\%$ \\
         & $10^5$ & $99\%$ & $99\%$ & $98\%$ & $98\%$ & $98\%$ & $98\%$ \\ \hline
        J-type shocks & Pre-shock density & Shock velocity (km/s)&  & & & & \\
         & (cm$^{-3}$) & 25 & 30 & 35 & 40 & 45 & 50 \\ \hline
         & $10^4$ & $0.2\%$ & $0.4\%$ & $0.7\%$ & $1.0\%$ & $1.1-1.4\%$ & $1.4-1.9\%$ \\
         & $10^5$ & $0.06\%$ & $0.25\%$ & $0.50\%$ & $0.7\%$ & $0.9\%$ & $1.0-1.2\%$ \\ \hline
         Comparison factor & Pre-shock density & Shock velocity (km/s)&  & & & & \\
         & (cm$^{-3}$) & 25 & 30 & 35 & 40 & 45 & 50 \\ \hline
         & $10^4$ & $2.1 \times 10^{-3}$ & $4.1 \times 10^{-3}$ & $7.3 \times 10^{-3}$ & $1.0 \times 10^{-2}$ & $1.2 \times 10^{-2}$ & $1.5 \times 10^{-2}$\\
         & & & & & & $1.5 \times 10^{-2}$ & $2.0 \times 10^{-2}$ \\
         & $10^5$ & $6.1 \times 10^{-4}$ & $2.5 \times 10^{-3}$ & $5.1 \times 10^{-3}$ & $7.1 \times 10^{-3}$ & $9.2 \times 10^{-3}$ & $1.0 \times 10^{-2}$\\
         & & & & & & & $1.2 \times 10^{-2}$ \\ \hline
    \end{tabular}
    \label{tab:outofrange}
    \tablefoot{Percentages evaluated for C-type shocks are computed with the Paris-Durham Shock code and correspond to the maximum of gas-phase methanol abundance divided by the initial icy-mantle phase abundance. Data for J-type shocks comes from Table 1 in \cite{2009A&A...497..145G} and corresponds to the percentage of released carbon from vaporization and sputtering. It is used as a proxy of the upper limit of the methanol's desorption percentage. The magnetic field parameter for C-type shocks has been chosen to be $10\%$ of the shock velocity to ensure that the magnetosonic speed is high enough for the shock to exist.}
\end{table*}

As vaporization, shattering, and coagulation are not included in the Paris-Durham Shock code, we limited the shock velocity to 19 km/s, where their impacts are thought to be negligible \citep{2019A&A...622A.100G}. Their impacts become important for the desorption percentage of methanol at higher velocities in J-type shocks, where the ion-neutral drift does not occur \citep{2009A&A...497..145G}. However, even if it is out of the reach of our simulations, it is important to evaluate the behavior of shocks at higher velocities. This information is crucial to ensure that C-type shocks above the threshold of 12 km/s cannot be mistaken for J-type shocks that are outside our parameter space. For this purpose, Table \ref{tab:outofrange} compares the percentage of gas-phase methanol released from icy mantles in C-type shocks as computed by the Paris-Durham Shock code with the percentage of carbon released in the gas phase in J-type shocks including the effect of vaporization and sputtering computed in Table 1 in \cite{2009A&A...497..145G}. Both information are not identical. The percentage of gas-phase methanol released in C-type shocks was approximated by the ratio between the maximum methanol gas-phase abundance reached during the simulation and its initial surface abundance. Figure \ref{fig:DesorptionPercentage1} represents the average value between the shock front and the end of the shock. Although this choice is closer to observables, it means that this percentage is already impacted by gas-phase processes. For J-type shocks, the percentage of released carbon is used as a proxy, but it does not include the degradation of methanol once it has reached the gas phase. For this reason, and knowing that gas-phase methanol is mainly involved in destructive pathways, the percentage indicated for J-type shocks can be accepted as an upper limit of the percentage of desorption. This is sufficient to ensure a good comparison between shock types while limiting the influence of errors by computing the true desorption of methanol in J-type shocks.

The main indication from Table \ref{tab:outofrange} is that the desorption percentage of methanol in C-type shocks is always at least 30 times greater than its value for J-type shocks. Moreover, the desorption percentage for J-type shocks is always at least 10 times lower than the average desorption percentage of C-type shocks (Fig. \ref{fig:DesorptionPercentage1}) with a shock velocity higher than 15 km/s. Considering the values of J-type shocks as upper limits, the main conclusion is that C-type shocks with a velocity higher than 15 km/s can safely be diagnosed since their desorption percentages are significantly greater than those computed for J-type shocks for each velocity. Further investigations on the real desorption percentage of methanol in J-type shocks, decreasing the upper limits indicated in Table \ref{tab:outofrange}, would allow the identification of C-type shocks at lower velocities, until $\sim 12-13$ km/s where the threshold appears. Possible degeneracies of the desorption percentage at higher velocities can be safely ignored. C-type shocks at higher velocities are highly unlikely given the very high magnetic field needed to maintain their existence. Moreover, the high degradation of gas-phase methanol in high-velocity J-type shocks \citep{2014MNRAS.440.1844S} would keep a clear differentiation between the two shock types.

\subsection{Other shock-type tracers}
\label{sec:Other Shock-Type tracers}

As we mention above, we also considered H$_2$S and H$_2$CO as potential shock tracers as they have been observed in L1157 B2, but both have been disregarded. It has been found that, due to the exothermicity of its formation, H$_2$S directly desorbs from grains and is not stored as a reservoir \citep{2024MNRAS.531.1371B}. This deeply modifies its chemistry and its behavior in shocks. As the Paris-Durham Shock code does not consider chemical desorption, H$_2$S has been removed from our discussions. In our simulations, it has been found that H$_2$CO is efficiently formed in the gas phase in post-shock regions by hydrogenation of newly desorbed species. Its post-shock gas-phase abundance can even exceed its initial icy mantle phase abundance in some conditions. The assumption that the post-shock abundance is primarily determined by desorbed molecules is no longer valid, and formaldehyde has also been disregarded. However, methanol should not be the unique molecule able to identify shock types. It has been chosen because it is simple enough to be evaluated in the Paris-Durham Shock code chemical network and because it has been observed in L1157 B2. Future studies on more diversified COMs fulfilling the conditions presented in Sect. \ref{Meth:Identification} should be envisaged such as CH$_3$OCH$_3$ or HCOOH \citep{2019ApJ...881...32B}. As these other COMs are characterized by different binding energies, their desorption behavior should be shifted on the velocity axis. The use of multiple COMs in the future will lead to better constraints on the shock type, but also the shock velocity.

\section{Benchmarking and application to L1157 B2} \label{sec:testcase}

Our paper introduces a new methodology based on methanol desorption to identify the shock type in cases where the type is ambiguous. It only requires measurements of gas-phase methanol in the post-shock region and ice-phase methanol in the pre-shock region. Although the first is commonly available, the latter is rarer due to the lack of emission in the radio domain. Even if observations with the \textit{James Webb} Space Telescope will make this value more accessible in the future \citep{2024A&A...683A.124R}, with the potential to validate further our method, we still need to approximate it for many real cases. Two measures are considered pertinent in protostellar environments. Firstly, recent simulations of COMs formation in cold cores found that an abundance of the order of $10^{-8}$ for gas-phase methanol corresponds to a fractional abundance of the order of $3 \times 10^{-5}$ for methanol in the icy mantle-phase \citep{2025ApJS..277....8L}. This abundance in the gas-phase is in great agreement with the pre-shock fractional abundance of methanol observed in the gas-phase in L1157 B2 \citep{1997ApJ...487L..93B}. Secondly, infrared observations from the \textit{James Webb} Space Telescope measured the solid-phase fractional abundance of CH$_3$OH compared to icy water around $\sim$ 6\% near IRAS 2A, a young protostar \citep{2024A&A...683A.124R}. With the initial water fractional abundance of $1.03 \times 10^{-4}$ used by the Paris-Durham Shock code, it corresponds to a fractional abundance of methanol around $\sim 6 \times 10^{-6}$. However, this value would correspond to a desorption percentage higher than $100\%$ for various cases. This seems nonphysical in view of the behavior analyzed in this paper and previous studies \citep{2014MNRAS.440.1844S, 2019ApJ...881...32B} and the value should only be valid as a lower limit. This low value is not necessarily due to the observations in \cite{2024A&A...683A.124R}, but can be a consequence of the initial fractional abundance of water in the solid-phase used by the Paris-Durham Shock code.  We thus decided to keep the mean value of the abundances ($1.8 \times 10^{-5}$) as a first approximation of the pre-shock ice-phase methanol abundance in protostellar outflows. An uncertainty exists in this initial observed icy-mantle phase fractional abundance of methanol, and thus in the observed desorption percentage. Despite these uncertainties, we were able to develop a methodology for diagnosing shock type, which we present in the next section. We note that only the uncertainty of the observed initial abundance is important, since the simulated one is not dependent on the initial conditions (see Sect. \ref{sec:Robustness}).

\subsection{Benchmarking}

\begin{table*}[]
    \centering
    \caption{Identification of the shock type through methanol desorption for various shocks for which the shock type is known. }
    \begin{tabular}{l l l l l l l}
    \hline
        Shock name & Pre-shock density & Shock-type & $\chi_{post-shock} [CH_3OH]$ & Shock type & Shock type & Remarks \\
         & Shock velocity & from literature & & without velocity & with velocity & \\ \hline
        L1157 B1\tablefootmark{a} & $10^4-10^5$ cm$^{-3}$ & C-type & $9 \times 10^{-6}$ & C-type & C-type & $\%_{deso} = 50\%$ \\
         & 40 km/s & & & & & $\sigma_{min} = 1.3$ \\
        L1448-mm\tablefootmark{b} & $1 \times 10^5$ cm$^{-3}$ & C-type & $10^{-6}$ & Unknown & C-type & $\%_{deso} = 6\%$ \\
         & $10 - 15$ km/s & & & & & $\sigma_{min} = 1.4$ \\
        HH 7-11\tablefootmark{c} & $10^4$ cm$^{-3}$ & J-type & $7.4 \times 10^{-9}$ & Unknown & J-type & $\%_{deso} = 0.04\%$ \\
         & $40 - 50$ km/s & & & & & $\sigma_{min} = 12.5$\tablefootmark{e} \\
        W3 IRS5\tablefootmark{d} & $10^4$ cm$^{-3}$ & J-type & $6.2 \times 10^{-10}$ & Unknown & Unknown\tablefootmark{f} & $\%_{deso} = 3.4 \times 10^{-3}\%$ \\
         & 10 km/s & & & & & $\sigma_{min} = 1.0$\\ \hline
    \end{tabular}
    \tablefoot{
     The $\chi_{post-shock} [CH_3OH]$ indicates the observed fractional abundance of methanol in the post-shock medium. The shock type without velocity indicates an identification without any constraints on the velocity. The shock type with velocity means that we reduce the parameter space to shocks with a velocity included in $[V_s\,\text{-}\,5, V_s + 5]$, $V_s$ being the shock velocity found in the literature.

The type of shock is considered to be identified if all models with geometric standard deviations less than three times the standard deviation (Eq. \ref{geometricdeviation}) of the closest model are of the same type. Appendix \ref{AppendixB} contains the geometric deviation for all shocks marked here. \\
    \tablefoottext{a}{For L1157 B1, the pre-shock density of $10^5$ cm$^{-3}$ and the velocity were inferred in \cite{2011ApJ...740L...3V}. A pre-shock density of $10^4$ cm$^{-4}$ was found in \cite{2013ApJ...776...52C}. These pre-shock densities are in accordance with \cite{2013MNRAS.436..179B}. All these studies agree on the type of L1157 B1.  The fractional abundance of methanol is the mean value from \cite{2020A&A...635A..17C} and \cite{1997ApJ...487L..93B} as it is included in their uncertainties.}
    \tablefoottext{b}{L1448-mm is associated with several outflows, including a very powerful one \citep{1990A&A...231..174B}. In these outflows, one of them is thought to be a C-type shock with a density of $10^5$ cm$^{-3}$, and a velocity of $10$ km/s \citep{2004ApJ...603L..49J, 2005ApJ...627L.121J}. Another limit at $15$ km/s has been given in \cite{2010ApJ...717...58H}}. The methanol abundance was derived in \cite{2005ApJ...627L.121J}
    \tablefoottext{c}{The density, the velocity, and the shock type of HH 7-11 are discussed in \cite{2000ApJ...538..698M}. The fractional abundance of methanol is from \cite{2006A&A...449.1089V}.}
    \tablefoottext{d}{The shock velocity, the pre-shock density, and the shock type of W3 IRS5 represent the best match in \cite{2003Ap&SS.284.1143A}. The methanol abundance was measured in \cite{1997A&AS..124..205H}.}
    \tablefoottext{e}{$\sigma_{min}$ is identified in the reduced parameter space with $V_s \in [35, 50]$ km/s.}
    \tablefoottext{f}{W3 IRS5 equally matches a C-type shock and a J-type shock model with $\sigma_{min} = 1.0$. The safe identification as a J-type shock can only be done with a strong constraint on the pre-shock density at $10^4$ cm$^{-3}$, with $\sigma_{min} = 1.1$.}
    }
    \label{tab:benchmarking}
\end{table*}

To validate our method before the shock-type diagnosis of L1157 B2, we applied it to various shocked environments for which the shock type is known and for which the post-shock gas-phase fractional abundance of methanol has been measured. The comparison was made using the geometrical standard deviation $\sigma_g$ (Eq. \ref{geometricdeviation}) between the observed desorption percentage value and our predictions to choose which model best matches the studied shock. A geometrical deviation is needed to distinguish the shock models with sufficient contrast. An arithmetic deviation would not efficiently compare models, given that the simulated desorption percentage can quickly become negligible compared to the observed value, consequently leading to the same arithmetic error for almost all shock models. Results are found in Table \ref{tab:benchmarking}. An identification was considered unambiguous if all models with geometric standard deviations less than three times the standard deviation of the best model are of the same type. The limit at 3-sigma means that the observed desorption percentage can be multiplied or divided by 3 without modifying the result of the identification. Like this, we expect to keep safe identifications despite uncertainties on the observed desorption percentage. In Table \ref{tab:benchmarking}, the identification without velocity did not assume any prior knowledge of the velocity of the analyzed shock. In this first identification, the shock velocity was thus comprised between 5 km/s and 50 km/s. For this, we scanned all the desorption percentages given in Fig. \ref{fig:DesorptionPercentage1} and Table \ref{tab:outofrange}. In contrast, the identification with velocity reduced the parameter space to the velocities included between $[V_s\,\text{-}\,5, V_s + 5]$, $V_s$ being the shock velocity found in the literature. This enables us to evaluate how much a shock should be constrained to be identified through our method:

\begin{equation} \label{geometricdeviation}
    \sigma_g = exp \left ( \sqrt{ \left (ln \frac{\%_{deso}^{computed}}{\%_{deso}^{ref}} \right )^2} \right ).
\end{equation}

Some conclusions follow from the results in Table \ref{tab:benchmarking} and the behavior of the desorption percentage. A high desorption percentage ($\%_{deso} > 30$) can only be explained by C-type shocks with shock velocity higher than 15 km/s (Fig. \ref{fig:DesorptionPercentage1}). Shocks presenting a high desorption percentage of methanol, such as L1157 B1, are thus always identified as C-type shocks, without needing any further constraints on the velocity. When the desorption percentage decreases, as for L1448-mm, a degeneracy appears between C-type shocks with less efficient sputtering ($V_s \in [12, 14]$ km/s) and high-velocity J-type shocks ($V_s \geq 25$ km/s). In this context, the method requires constraints on the velocity for a reliable identification of the shock type. However, as explained previously, the desorption percentage for high-velocity J-type shocks, as presented in Table \ref{tab:outofrange}, is a very pessimistic upper limit given that it does not include the degradation of gas-phase methanol, even if it is known to occur \citep{2014MNRAS.440.1844S}. A more realistic calculation of the percentage of desorption in high-velocity J-type shocks should enable a better characterization of C-type shocks, without any constraints on velocity.

A differentiation between C-type shocks and J-type shocks is only feasible in the domain of the parameter space where the comparison factor is far enough from 1. More specifically, the comparison factor has to be greater than 3 (or lower than $\frac{1}{3}$) to enable identification using the 3-sigma test. Identification of a J-type shock is thus possible only for shocks for which the velocity is greater than 10 km/s. In addition, further constraints on the velocity are required to differentiate C-type shocks at intermediate velocities ($V_s \in [12; 14]$ km/s) and high-velocity J-type shocks ($V_s \geq 25$ km/s). Low-velocity J-type shocks ($V_s$ possibly less than 10 km/s), such as W3 IRS5, cannot be identified only using constraints on the shock velocity. As discussed in Table \ref{tab:benchmarking}, strong constraints on the pre-shock density can lead to a nonambiguous identification. However, such constraints are difficult to obtain. In the future, better computations of the desorption percentage, coupled with observations of methanol abundances and shock conditions (velocity and pre-shock density) with smaller uncertainties, should also lead to easier identifications. The last conclusion drawn from Table \ref{tab:benchmarking} is that the method does not seem to create false identification, and therefore can furnish a reliable identification of L1157 B2.

\subsection{Identification of L1157 B2 shock type}

\begin{figure*}
    \centering
    \includegraphics[width=1.2\linewidth]{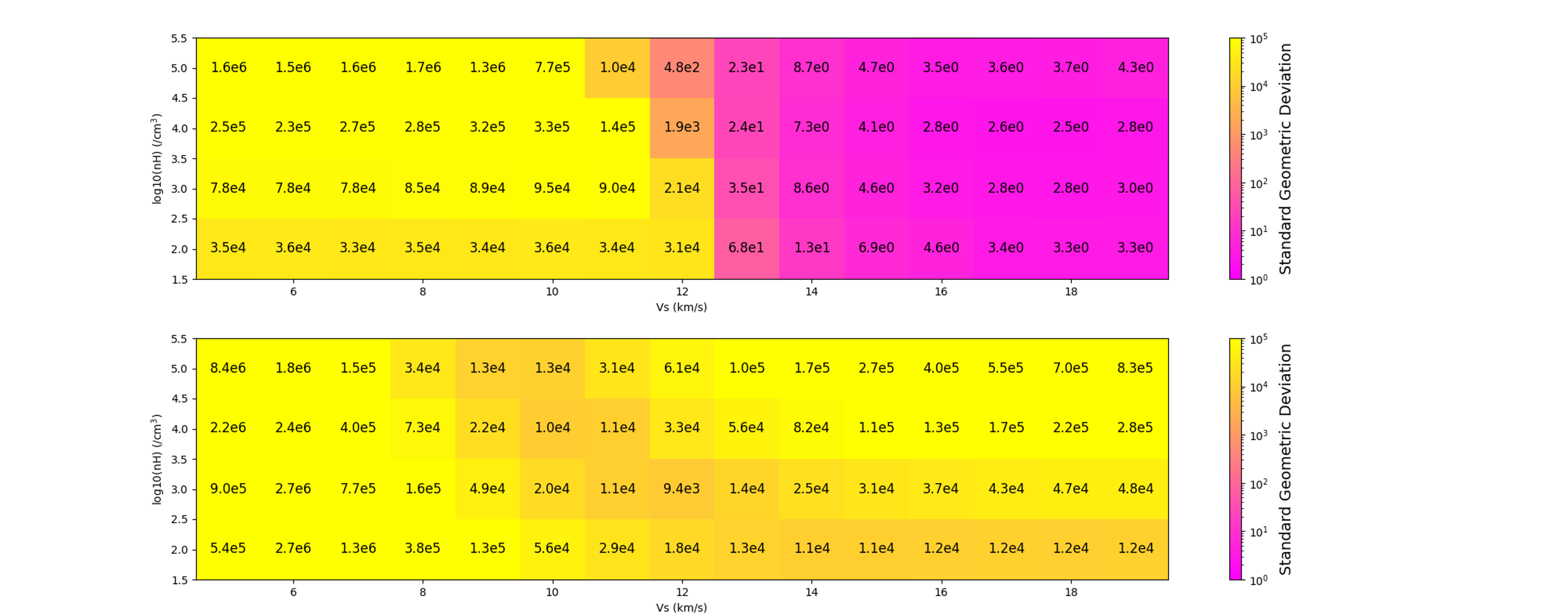}
    \caption{Geometric deviation between simulations and L1157 B2 for C-type shocks (top figure) and J-type shocks (bottom figure). The x-axis represents the shock velocity ($V_s$) and the y-axis the pre-shock density ($nH$).}
    \label{fig:GeometricDeviation}
\end{figure*}

The post-shock gas-phase methanol abundance of L1157 B2 is $2.2 \times 10^{-5}$ \citep{1997ApJ...487L..93B}. With a pre-shock ice-phase abundance of $1.8 \times 10^{-5}$, it corresponds to a desorption percentage of $122\%$. As this desorption percentage is unrealistic, we decided to limit it to $100\%$ to keep it in a physically valid range, although this value is certainly overestimating the actual value. Destructive pathways encountered by methanol prevent the post-shock gas-phase methanol abundance from recovering its pre-shock icy-mantle phase value. The very high desorption percentage obtained in this paper is the consequence of the low initial grain phase methanol abundance we used as an estimator. A better estimation of the percentage will only be possible with further observations of ice-phase methanol near protostellar environments. However, tests with a desorption percentage of $122\%$ would yield the same identification, and the method has been proven to be robust to a wide range of desorption percentages for shocks with high desorption ability. The application of the geometric standard deviation (Eq. \ref{geometricdeviation}) can be found in Fig. \ref{fig:GeometricDeviation}.

Without any constraints on the velocity, the minimum geometric standard deviation is $\sigma_{min} = 1.01$, and corresponds to the model of a C-type shock with a pre-shock density of $10^5$ cm$^{-3}$ and a shock velocity of 25 km/s. However, values for shocks in Table \ref{tab:outofrange} do not correspond to a desorption percentage of methanol averaged between the shock front and the end of the shock as in Fig. \ref{fig:DesorptionPercentage1}, but to the desorption percentage using the maximum gas-phase methanol abundance reached during the simulation. They are thus overestimated. A comparison inside the parameter space (Fig. \ref{fig:DesorptionPercentage1}) indicates that the best model ($\sigma_{min} = 2.5$) is a C-type shock with high enough velocity ($V_s = 18$ km/s) to ensure efficient sputtering of methanol. With this geometric standard deviation, all models in the parameter space with a geometric standard deviation below $3 \sigma_{min}$ are C-type shocks with $V_s \geq 14$ km/s. L1157 B2 can thus be safely diagnosed as a C-type shock. This identification does not include any constraints on the velocity, but constraining the parameter space around 10 km/s \citep{2016MNRAS.462.2203G} leads to the same identification. However, the original constraint on the velocity was not robust. \cite{2016MNRAS.462.2203G} did not focus on the characterization of L1157 B2. The shock velocity of 10 km/s only represents the best match between observations and simulations, but 5 km/s separates the velocities in the set of models, and the models at 15 km/s are never tested at a pre-shock density of $10^3$ cm$^{-3}$. The geometric standard deviation can be used to better constrain the shock velocity on L1157 B2. Given that L1157 B2 presents a very high desorption percentage, Fig. \ref{fig:GeometricDeviation} indicates that it can only be explained by C-type with a shock velocity higher than 13 km/s using the limit at $\sigma < 3 \times \sigma_{min} = 7.5$. We thus put further constraints on the shock velocity. Our method does not, however, provide any new constraints on the pre-shock density.

\subsection{Impact of the FUV radiation field}

In the previous sections, we assumed that shocks operated without the influence of any external FUV radiation. This is justified by the formation of these shocks near class 0 protostellar cores. Class 0 are known to be located in embedded systems, given that their masses are still comparable to the mass of their environment \citep{1994ASPC...65..197B}, where the FUV radiation field is strongly weakened. Moreover, as previously mentioned for the test case, L1157 B2 is located near L1157 B1, a denser region that strongly attenuates the FUV radiative field \citep{2014A&A...565A..64P}. A second proof that L1157 B2 is a non-irradiated C-type shock is brought by the applicability of the method. Simulations of irradiated shocks have been performed for C-type, J-type, CJ-type, and C$^*$-type shocks using the Paris-Durham Shock code. It has been found that an external FUV radiation field of $0.1$ Habing units was already enough to strongly desorb the icy mantle of dust grains in the pre-shocked medium. This led to a very small methanol abundance, which was destroyed into smaller molecules. Only a small quantity (of the order of $6 \times 10^{-16}$) remained due to gas-phase formation from CH$_5$O$^+$. The presence of a high abundance of gas-phase methanol in the post-shock region of L1157 B2, which is representative of a relatively large initial amount trapped in icy mantles, indicates that L1157 B2 is a non-irradiated C-type shock. The only uncertainty could come from HH 7-11, which is thought to originate from a star being at the intersection of class 0 and class 1. However, Herbig-Haro objects trace the interface between outflows and circumstellar materials \citep{2000ApJ...538..698M}, and class 1 protostellar cores are still too cold to be strong emitters of FUV photons. FUV photons should come from the shock itself, but not as an external FUV irradiation field.

\section{Conclusions}
\label{sec:conclusions}

We proposed a new criterion for determining the shock type of non-irradiated low-velocity outflows. Through simulations of non-irradiated J-type and C-type shocks, we found that the post-shock abundance of gas-phase methanol – and more specifically, its ratio compared to icy mantle initial abundance – is dependent on the shock type and presents a good contrast for shocks having a shock velocity higher than 15 km/s. Consequently, it gives information on the shock conditions within non-irradiated environments. Through our analysis, we confirmed through benchmarking the shock type of already characterized shocks and succeeded in diagnosing L1157 B2 as a non-irradiated C-type shock. The information on its shock type was still uncertain, given that C-type and J-type shocks have many similarities \cite{2020A&A...634A..17J}. For the velocity, our analysis fits with previous studies and suggests a lower limit on the shock velocity at 13 km/s. It is interesting to note that, contrary to previously reported discussions, molecular abundances can be used to diagnose shock types. Even if molecular abundances can be very condition-specific, it is possible to use them to define criteria that are not dependent on the initial conditions. This is actually the case for the desorption percentage. Diagnosis of astrophysical environments through astrochemistry is encouraged.

In the current scientific understanding of methanol in shocks, our analyses pursue and validate the work that has been initiated by other teams to understand its complex behavior. Moreover, it strongly complements the observations made by \cite{2014MNRAS.440.1844S} by rigorously comparing shock types at all velocities. As numerous pieces of evidence arise for relatively high-velocity C-type shocks, such as L1157 B1, or low-velocity J-type shocks, such as W3 IRS5, the shock type is no longer a good indicator of the velocity and vice versa. A good understanding of the shock type influences on chemicals will thus be essential to determine the shock type, to explain chemistry in shocks, but also to constrain other physical properties that are at the basis of shock formation, such as the magnetic field.

Directly concerning the method proposed in this paper, three major improvements are expected in the future:

\begin{enumerate}
    \item Even if our method is as least model-dependent as possible, we still rely on the ability of the model to compute desorption processes. More rigorous methods including distributions of binding energy are necessary to increase the precision of the method. They seem even obligatory in the context of molecules with lower binding energy or wide distributions, such as in the case of NH$_3$ and its double-peak distribution \citep{Tinacci,2025A&A...698A.284G}. Furthermore, a separation of the icy mantle phase into an interior and an envelope could also modify the results. This would also be beneficial to differentiate cosmic-ray-induced desorption and photodesorption, with different penetration depths. The inclusion of shattering and vaporization will also be essential if denser shocks are considered.

    \item Our analysis considers that we know the initial icy mantle phase abundance of the species of interest. This can be a real drawback for species that are less constrained. Nevertheless, we expect the advent of more complex COM formation simulations and infrared observations such as the ones coming from the \textit{James Webb} Space Telescope to help in the diagnosis of shock types. Future observations with the \textit{James Webb} Space Telescope should make the application of the methanol desorption method to characterize shock type more accessible, with the potential to expand this approach to other targets.

    \item The method is based on methanol initially trapped in icy mantles. It thus only works for environments sufficiently shielded from the external FUV radiation field for this ice layer to exist. However, different types of shocks are also possible in an irradiated medium, and subsequent methods should be found to characterize them. Nevertheless, COMs desorption is well-suited for non-irradiated shocks, such as the ones found during the class 0 phase.
\end{enumerate}

Methanol was a good indicator for this study, but should not be considered as the only one. COMs that are mainly formed in icy mantles can have similar behaviors. As they have different binding energies, analysis based on observations of other COMs could lead to better constraints and smaller uncertainties. Investigations on COM behavior in shocks are encouraged.

\begin{acknowledgements}
Data treatment and plots were made using Numpy \citep{harris2020array} and Matplotlib \citep{Hunter:2007}. This research has made use of the Astrophysics Data System, funded by NASA under Cooperative Agreement 80NSSC21M00561. Computational resources have been provided by the Consortium des Équipements de Calcul Intensif (CÉCI), funded by the Fonds de la Recherche Scientifique de Belgique (F.R.S.-FNRS) under Grant No. 2.5020.11 and by the Walloon Region.
\end{acknowledgements}

\bibliographystyle{aa}
%\bibliography{aa53978-25}

\begin{appendix}

    \section{Analytical expression for the gas-phase methanol abundance} \label{AppendixA}

Being in a system where the concentration of chemicals is influenced by a compression rate $g(t) = - \dot V / V$, we write the chemical evolution of the system in the comoving reference density frame. The hydrogen density $nH$ and the concentration of species $[x]$, which is influenced by reaction rates $\sum_i k_i (t) [y_i] (t)$, evolve following Eq. \ref{eq:nH evolution} and \ref{eq:x evolution}:

\begin{equation} \label{eq:nH evolution}
    \frac{dnH(t)}{dt} = \dot{(\frac{N_H}{V})}(t) = - \frac{\dot V(t)}{V(t)} nH(t) = g(t) \times nH(t) ,
\end{equation}

\begin{equation} \label{eq:x evolution}
    \frac{d[x](t)}{dt} = \sum_i k_i(t) [y_i](t) + g(t) \times [x](t).
\end{equation}

The equation in the comoving reference density frame is thus given by Eq. \ref{eq:comoving}

\begin{align}
    \frac{d\chi[x]}{dt} &= \, \frac{d ([x] / nH)(t)}{dt} = \frac{\dot{[x]}(t) \times  nH(t) - [x](t) \times \dot{nH}(t)}{nH(t)^2} \\
     &= \frac{(\sum_i k_i [y_i] + g \times [x]) \times nH - g \times nH \times [x]}{nH^2} \\
     &= \sum_i k_i(t) \frac{[y_i](t)}{nH(t)} = \sum_i k_i(t) \times \chi [y_i](t), \label{eq:comoving}
\end{align}

\noindent where time-dependences have been abridged for clarity. The use of the comoving reference density always hold for J-type shock simulations where there is a full coupling between ions and neutrals. While some deviations occur near the shock front in C-type shock simulations, the approximation still holds when we average on the postshock region, as it is the case for the desorption percentage written in Eq. \ref{desorptionpercentage}. To obtain the time-dependent evolution of gas-phase methanol fractional abundance, we write the system of coupled PDE as a homogenous differential system as described in Eq. \ref{differentsystem}: 

\begin{dmath} \label{differentsystem}
    \frac{d}{dt} \begin{bmatrix}
        \chi[CH_3OH^*](t) \\
        \chi[CH_3OH](t)
    \end{bmatrix} = \begin{bmatrix}
        -k_{deso}(t) & k_{ads}(t) \\
        k_{deso}(t) & -(k_{ads}(t) + k_{dest}(t))
    \end{bmatrix} \cdot \begin{bmatrix}
        \chi[CH_3OH^*](t) \\
        \chi[CH_3OH](t)
    \end{bmatrix}
\end{dmath}

\noindent with the initial conditions (Eq. \ref{initialconditions}):

\begin{equation} \label{initialconditions}
    \chi[CH_3OH^*](t=t_0) = \chi[CH_3OH^*]_0
\end{equation}
\begin{equation}
    \chi[CH_3OH](t=t_0) = 0.
\end{equation}

In this system we have included every term that is not methanol concentration-dependant in the k constants. In other words, every rate reaction described by its rate constant ($k'_i$ - Eq. \ref{firstform}) is now described by a global constant ($k_i$ - Eq. \ref{secondform}):

\begin{equation} \label{firstform}
    \frac{d\chi[CH_3OH]}{dt} = - k'_i \cdot \chi[Reactant] \cdot \chi[CH_3OH]
\end{equation}

\begin{equation} \label{secondform}
    \frac{d\chi[CH_3OH]}{dt} = - k_i \cdot \chi[CH_3OH].
\end{equation}

From Cauchy’s theorem, the differential system has a single solution. Using Duhamel’s formula, its solution is of the form of Eq. \ref{GlobalEquation},

\begin{equation}\label{GlobalEquation}
    \begin{bmatrix}
        \chi[CH_3OH^*](t) \\
        \chi[CH_3OH](t)
    \end{bmatrix} = \begin{bmatrix}
        R_{11}(t) & R_{12}(t) \\
        R_{21}(t) & R_{22}(t)
    \end{bmatrix} \cdot \begin{bmatrix}
        \chi[CH_3OH^*]_0 \\
        0
    \end{bmatrix},
\end{equation}

\noindent where R(t) is the resolving matrix of the homogenous system. We can derive an analytical expression for its components in the quasi-static approximation assuming that the $k$ constants slowly vary compared to the evolution of the methanol fractional abundance. This assumption is unrealistic in shock conditions but necessary to obtain an analytical expression. However, as R(t) is considered to not depend on the methanol concentrations, we already know that our system is linear. It means that the basic approximations, i.e., (i) gas-phase pathways leading to methanol are negligible and (ii) $k$ constants are not methanol concentration-dependent, are enough to get a desorption percentage that is not dependent on initial conditions. Under quasi-static approximation, R(t) can be written as the product of $P(t)$ and $P^{-1}(t)$, the eigenvectors matrix and its inverse, and the exponential of the time-integral of $\Lambda(t)$ (the diagonal eigenvalues matrix) as in Eq. \ref{Requation}:

\begin{equation} \label{Requation}
    R(t) = P(t) \cdot exp \left ( \int_{t_0}^{t} \Lambda(\tau) \; d\tau \right ) \cdot P^{-1}(t).
\end{equation}

Computing the eigenvalues of the system, we obtain the expression given in Eq. \ref{eigenvalues}. From these, the eigenvectors matrixes $P(t)$ and $P^{-1}(t)$ can be written as in Eq. \ref{Pmatrix}, \ref{Pimatrix}:

\begin{equation} \label{Pmatrix}
    P(t) = \begin{bmatrix}
        1 & 1 \\
        \frac{k_{deso}(t) + \lambda_1(t)}{k_{ads}(t)} & \frac{k_{deso}(t) + \lambda_2(t)}{k_{ads}(t)}
    \end{bmatrix}
\end{equation}

\begin{equation} \label{Pimatrix}
    P^{-1}(t) = \frac{k_{ads}(t)}{\lambda_2(t) - \lambda_1(t)} \cdot \begin{bmatrix}
        \frac{k_{deso}(t) + \lambda_2(t)}{k_{ads}(t)} & -1 \\
        - \frac{k_{deso}(t) + \lambda_1(t)}{k_{ads}(t)} & 1
    \end{bmatrix}.
\end{equation}

As we are only interested in gas-phase methanol evolution, it comes from Eq. \ref{GlobalEquation} that only $R_{21}(t)$ needs to be evaluated. Making use of the definition of $\Phi(t, t_0)$ (Eq. \ref{phi}), it takes the form of Eq. \ref{finalequation}:

\begin{align}\label{finalequation}
    R_{21}(t) &= P_{21}(t) \cdot e^{\Phi_1(t, t_0)} \cdot P_{11}^{-1}(t) + P_{22}(t) \cdot e^{\Phi_2(t, t_0)} \cdot P_{21}(t) \\
     &= \frac{(k_{deso}(t) + \lambda_1(t)) \cdot (k_{deso}(t) + \lambda_2(t))}{(\lambda_2(t) - \lambda_1(t)) \cdot k_{ads}(t)} \cdot (e^{\Phi_1(t, t_0)} - e^{\Phi_2(t, t_0)}) \\
     &= \%_{deso}(t)
\end{align}

\section{Geometric deviation for the benchmarking} \label{AppendixB}

\begin{figure*}
    \centering
    \includegraphics[width=1.2\linewidth]{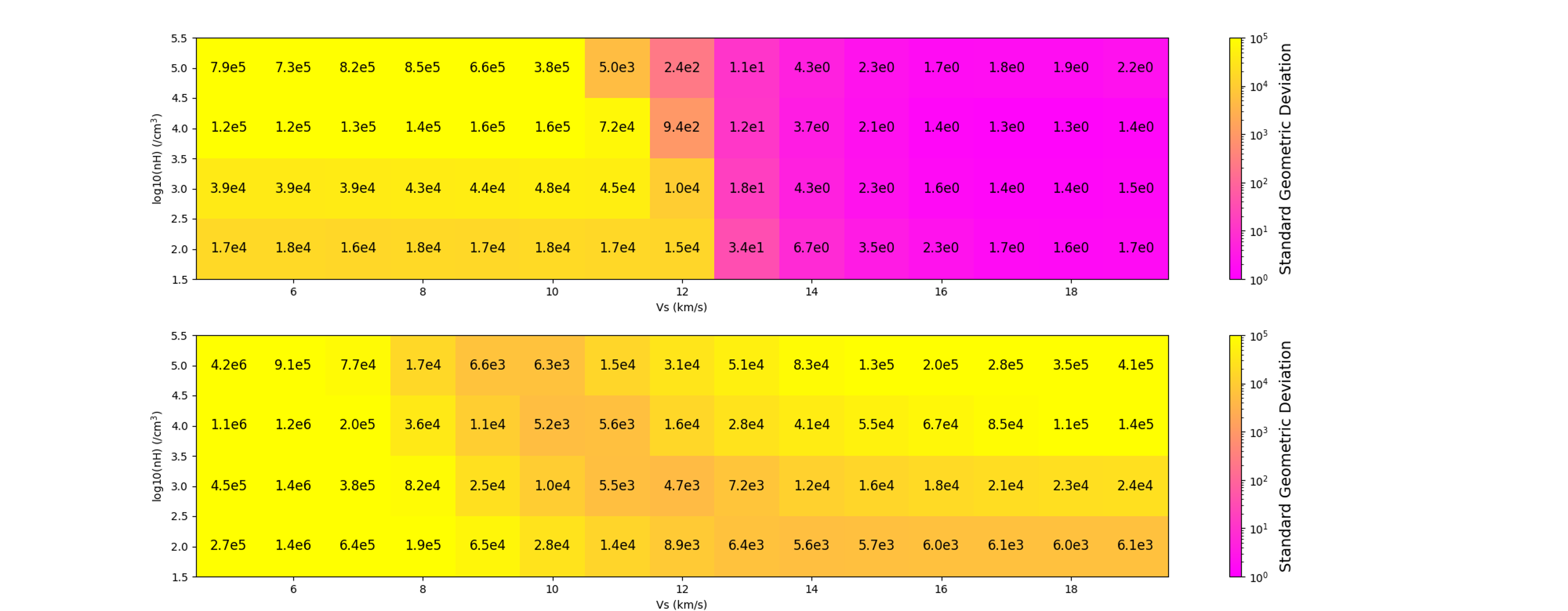}
    \caption{Geometric deviation between simulations and L1157 B1 for C-type shocks (top) and J-type shocks (bottom). The x-axis represents the shock velocity ($V_s$) and the y-axis the pre-shock density ($nH$).}
    \label{fig:GeometricDeviationL1157B1}
\end{figure*}

\begin{table*}[]
    \centering
    \caption{Geometric deviation for L1157 B1.}
    \begin{tabular}{l l l l l l l l}
    \hline
        C-type shocks & Pre-shock density & Shock velocity (km/s)&  & & & & \\
         & (cm$^{-3}$) & 25 & 30 & 35 & 40 & 45 & 50 \\ \hline
         & $10^4$ & $1.9$ & $1.9$ & $1.9$ & $1.9$ & $1.9$ & $1.9$ \\
         & $10^5$ & $2.0$ & $2.0$ & $2.0$ & $2.0$ & $2.0$ & $2.0$ \\ \hline
        J-type shocks & Pre-shock density & Shock velocity (km/s)&  & & & & \\
         & (cm$^{-3}$) & 25 & 30 & 35 & 40 & 45 & 50 \\ \hline
         & $10^4$ & $250$ & $125$ & $71$ & $50$ & $45 - 36 $ & $36-26$ \\
         & $10^5$ & $833$ & $200$ & $100$ & $71$ & $56$ & $50-42$ \\ \hline
    \end{tabular}
    \label{tab:L1157 B1}
    \tablefoot{Geometric deviation between simulations and L1157 B1 in C-type shocks and J-type shocks using the data given in Table \ref{tab:outofrange}.}
\end{table*}

\begin{figure*}
    \centering
    \includegraphics[width=1.2\linewidth]{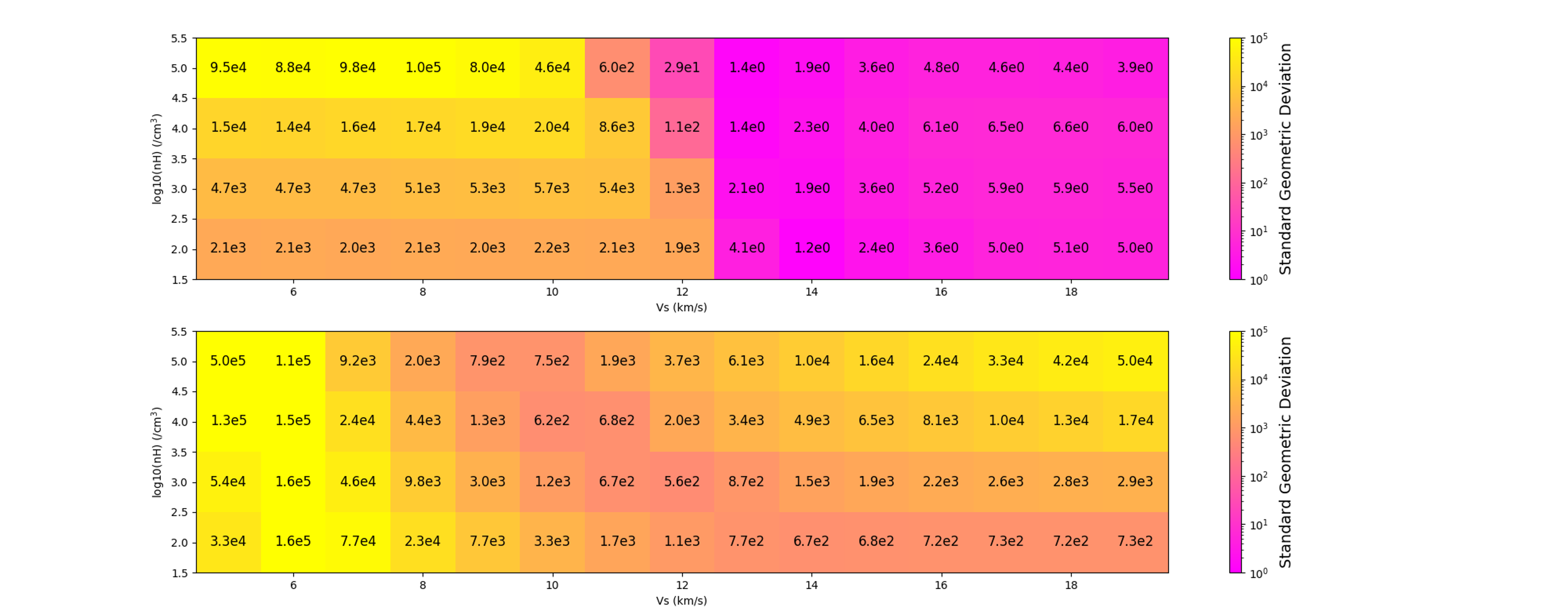}
    \caption{Geometric deviation between simulations and L1448-mm for C-type shocks (top) and J-type shocks (bottom). The x-axis represents the shock velocity ($V_s$) and the y-axis the pre-shock density ($nH$).}
    \label{fig:GeometricDeviationL1448mm}
\end{figure*}

\begin{table*}[]
    \centering
    \caption{Geometric deviation for L1448-mm.}
    \begin{tabular}{l l l l l l l l}
    \hline
        C-type shocks & Pre-shock density & Shock velocity (km/s)&  & & & & \\
         & (cm$^{-3}$) & 25 & 30 & 35 & 40 & 45 & 50 \\ \hline
         & $10^4$ & $16$ & $16$ & $16$ & $16$ & $16$ & $16$ \\
         & $10^5$ & $16$ & $16$ & $16$ & $16$ & $16$ & $16$ \\ \hline
        J-type shocks & Pre-shock density & Shock velocity (km/s)&  & & & & \\
         & (cm$^{-3}$) & 25 & 30 & 35 & 40 & 45 & 50 \\ \hline
         & $10^4$ & $30$ & $15$ & $8.6$ & $6.0$ & $5.4-4.3$ & $4.3-3.2$ \\
         & $10^5$ & $100$ & $24$ & $12$ & $8.6$ & $6.7$ & $6.0-5.0$ \\ \hline
    \end{tabular}
    \label{tab:L1448 mm}
    \tablefoot{Geometric deviation between simulations and L1448-mm in C-type shocks and J-type shocks using the data given in Table \ref{tab:outofrange}.}
\end{table*}

\begin{figure*}
    \centering
    \includegraphics[width=1.2\linewidth]{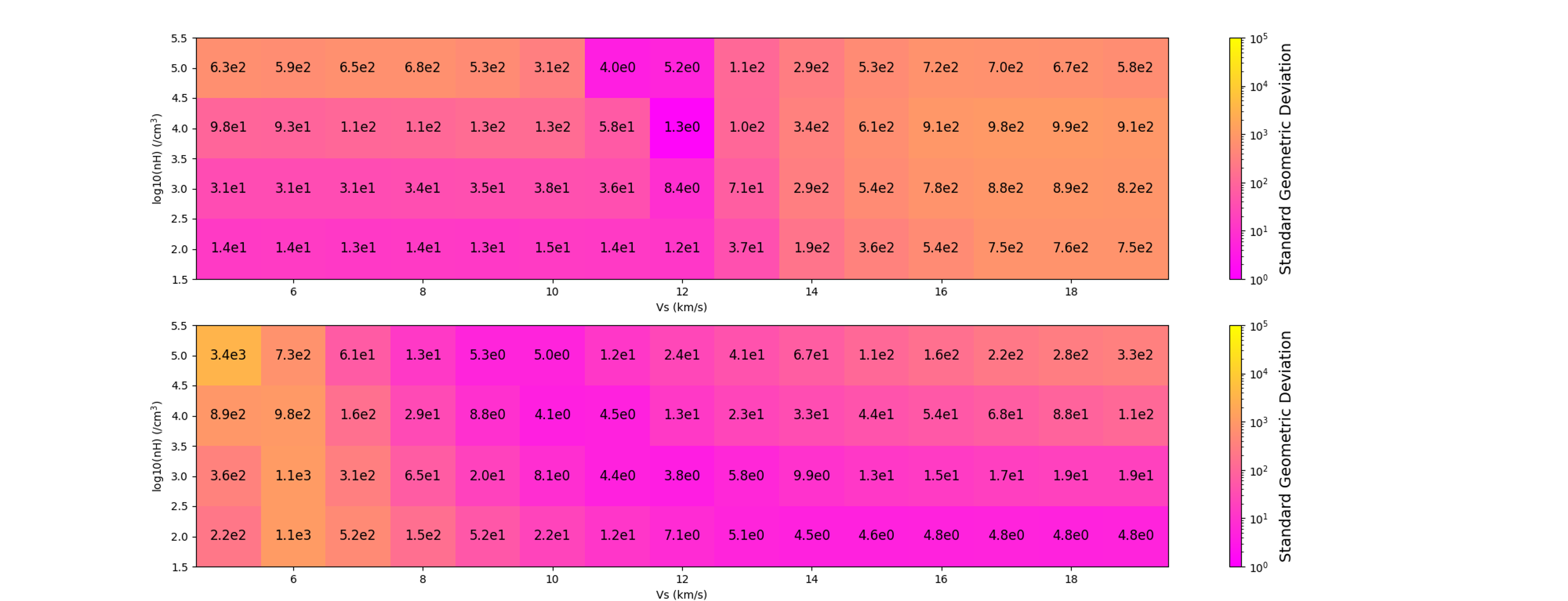}
    \caption{Geometric deviation between simulations and HH 7-11 for C-type shocks (top) and J-type shocks (bottom). The x-axis represents the shock velocity ($V_s$) and the y-axis the pre-shock density ($nH$).}
    \label{fig:GeometricDeviationHH711}
\end{figure*}

\begin{table*}[]
    \centering
    \caption{Geometric deviation for HH 7-11.}
    \begin{tabular}{l l l l l l l l}
    \hline
        C-type shocks & Pre-shock density & Shock velocity (km/s)&  & & & & \\
         & (cm$^{-3}$) & 25 & 30 & 35 & 40 & 45 & 50 \\ \hline
         & $10^4$ & $2425$ & $2425$ & $2400$ & $2375$ & $2375$ & $2375$ \\
         & $10^5$ & $2475$ & $2475$ & $2450$ & $2450$ & $2450$ & $2450$ \\ \hline
        J-type shocks & Pre-shock density & Shock velocity (km/s)&  & & & & \\
         & (cm$^{-3}$) & 25 & 30 & 35 & 40 & 45 & 50 \\ \hline
         & $10^4$ & $5.0$ & $10$ & $18$ & $25$ & $28-35$ & $35-48$ \\
         & $10^5$ & $1.5$ & $6.2$ & $12.5$ & $18$ & $22$ & $25-30$ \\ \hline
    \end{tabular}
    \label{tab:HH 7 11}
    \tablefoot{Geometric deviation between simulations and HH 7-11 in C-type shocks and J-type shocks using the data given in Table \ref{tab:outofrange}.}
\end{table*}

\begin{figure*}[]
    \centering
    \includegraphics[width=1.2\linewidth]{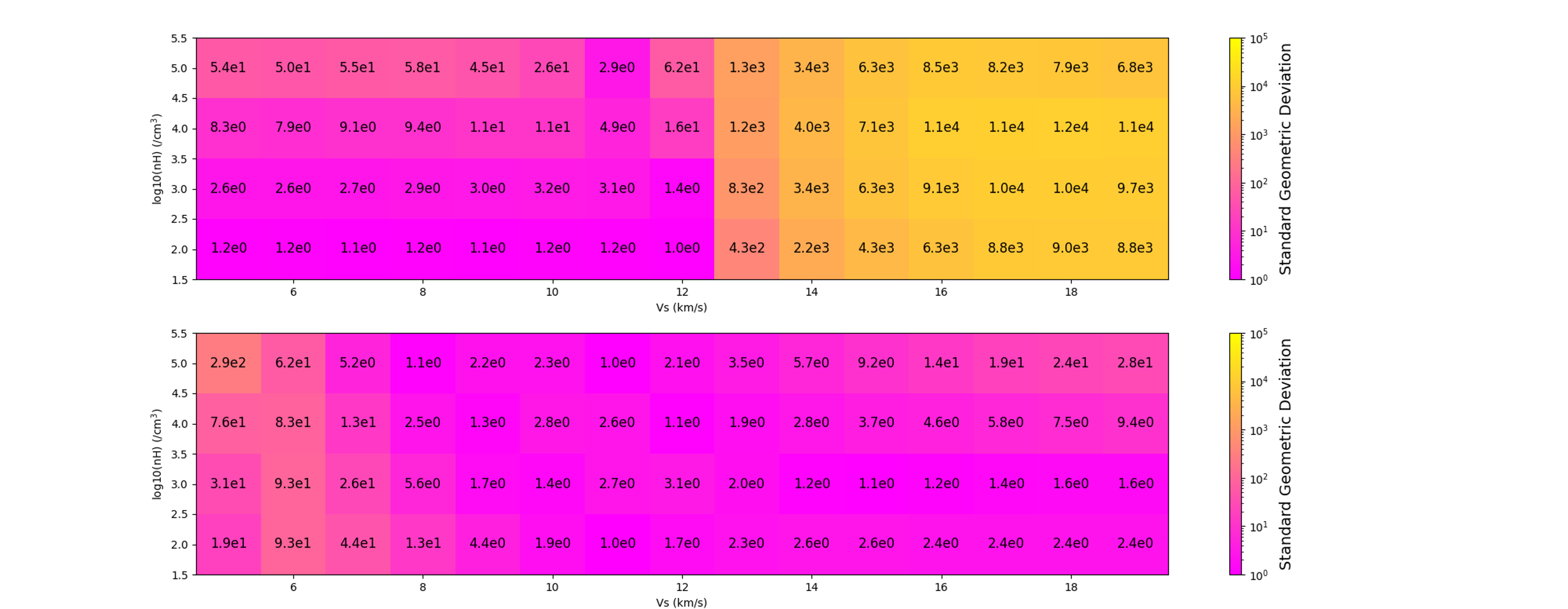}
    \caption{Geometric deviation between simulations and W3 IRS5 for C-type shocks (top) and J-type shocks (bottom). The x-axis represents the shock velocity ($V_s$) and the y-axis the pre-shock density ($nH$).}
    \label{fig:GeometricDeviationW3IRS5}
\end{figure*}

\begin{table*}[]
    \centering
    \caption{Geometric deviation fr W3 IRS5 in C-type shocks.}
    \begin{tabular}{l l l l l l l l}
    \hline
        C-type shocks & Pre-shock density & Shock velocity (km/s)&  & & & & \\
         & (cm$^{-3}$) & 25 & 30 & 35 & 40 & 45 & 50 \\ \hline
         & $10^4$ & $2.8 \times 10^3$ & $2.8 \times 10^3$ & $2.8 \times 10^3$ & $2.8 \times 10^3$ & $2.8 \times 10^3$ & $2.8 \times 10^3$ \\
         & $10^5$ & $2.9 \times 10^3$ & $2.9 \times 10^3$ & $2.9 \times 10^3$ & $2.9 \times 10^3$ & $2.9 \times 10^3$ & $2.9 \times 10^3$ \\ \hline
        J-type shocks & Pre-shock density & Shock velocity (km/s)&  & & & & \\
         & (cm$^{-3}$) & 25 & 30 & 35 & 40 & 45 & 50 \\ \hline
         & $10^4$ & $59$ & $118$ & $206$ & $294$ & $324-412$ & $412-559$ \\
         & $10^5$ & $18$ & $74$ & $147$ & $206$ & $265$ & $294-353$ \\ \hline
    \end{tabular}
    \label{tab:W3 IRS5}
    \tablefoot{Geometric deviation between simulations and W3 IRS5 in C-type shocks and J-type shocks using the data given in Table \ref{tab:outofrange}.}
\end{table*}

\end{appendix}

\end{document}